%% file: main.tex
\documentclass[acmsmall,screen,nonacm]{acmart}

\usepackage{packages}
\ifPDFTeX\pdfdecimaldigits=5\fi
\microtypesetup{expansion=false}
\usepackage{commands}
\input{figures/f258-table-style}
\setcopyright{none}
\acmDOI{}
\acmISBN{}
\acmYear{2026}
\togglefalse{showcomments}
\renewcommand{\inlinecomment}[3]{}

\begin{document}

\title[Beyond Compilation Feedback]{Beyond Compilation Feedback: Refining Vulnerability Checkers with Program-Analysis Evidence}

\author{Chengzhi Yi}
\affiliation{\institution{Huazhong University of Science and Technology}\country{China}}
\email{yimou@hust.edu.cn}

\author{Shenao Wang}
\affiliation{\institution{Huazhong University of Science and Technology}\country{China}}
\email{shenaowang@hust.edu.cn}

\author{Yanjie Zhao}
\affiliation{\institution{Huazhong University of Science and Technology}\country{China}}
\email{yanjie\_zhao@hust.edu.cn}

\author{Xiao Cheng}
\affiliation{\institution{Macquarie University}\country{Australia}}
\email{jumormt@gmail.com}

\author{Haoyu Wang}
\affiliation{\institution{Huazhong University of Science and Technology}\country{China}}
\email{haoyuwang@hust.edu.cn}
\renewcommand{\shortauthors}{Yi et al.}

\begin{abstract}
\input{sections/00-abstract}
\end{abstract}
\maketitle


\section{Introduction}\label{sec:intro}
\input{sections/01-introduction}

\section{Background}\label{sec:background}
\input{sections/02-background}

\section{Motivating Example}\label{sec:motivating}
\input{sections/03-motivating-example}

\FloatBarrier
\section{Approach}\label{sec:approach}
\input{sections/03-approach}

\FloatBarrier
\section{Evaluation}\label{sec:evaluation}
\input{sections/04-evaluation}

\FloatBarrier
\section{Results}\label{sec:results}
\input{sections/05-results}

\section{Discussion}\label{sec:discussion}
\input{sections/06-discussion}

\section{Threats to Validity}\label{sec:threats}
\input{sections/07-threats}

\section{Related Work}\label{sec:related}
\input{sections/08-related}

\section{Conclusion}\label{sec:conclusion}
\input{sections/09-conclusion}


\begingroup
\interlinepenalty=10000
\bibliographystyle{ACM-Reference-Format}
\bibliography{references}
\endgroup

\end{document}

%% file: sections/00-abstract.tex
Recent work shows that LLMs can synthesize static-analysis checkers from security patches, but a compilable checker is not necessarily a correct one.
A patch reveals what changed, not why the change is safe, so generated checkers often match only the surface edit and miss the deeper semantic condition.
Existing refinement loops feed back compilation errors and validation reports, yet these signals describe symptoms without explaining what the checker misunderstands about the code.

We observe that the analysis framework itself already computes the information needed to close this gap.
Control-flow graphs, symbolic values, memory regions and checker callbacks are available inside the analyzer but unused by current refinement strategies.
\toolname is a plug-in refinement layer that extracts patch-relevant analyzer evidence and assembles it with source context and validation feedback into a structured, provenance-labelled bundle.
An iterative editing loop refines the checker on both the vulnerable and fixed revisions, guarded by a retention gate that accepts warning recovery or false-positive reduction while preventing over-pruning.

We evaluate \toolname on 39 KNighter-derived CSA checkers from Linux kernel fixes.
With analyzer evidence, refinement reaches 23 warning-positive vulnerable revisions (vs.\ 15 for KNighter's loop) and raises version-level $F_1$ from $0.448$ to $0.575$.
An ablation removing internal evidence recovers fewer silent cases and produces different checker edits, indicating that the two evidence configurations address complementary recognition gaps.
Additional experiments cover three editor models on a 12-subject subset, a CodeQL query refinement, and controlled source-form variations.
A qualitative follow-up on all 11 non-tied comparisons traces each warning-count change to a specific checker repair, such as recognizing a lowered callback or unwrapping an expression cast.

%% file: sections/01-introduction.tex
Static analysis is a cornerstone of software security assurance for systems code.
Tools such as CodeQL~\cite{codeqlDocs} and Clang Static Analyzer (CSA)~\cite{clangsaDocs} detect vulnerabilities by checking source code against predefined rules without executing it~\cite{engler2001bugs,bessey2010coverity,calcagno2015infer,sadowski2015tricorder}.
The effectiveness of these tools depends on the quality of their vulnerability checkers, each of which must encode the semantic conditions that distinguish a real bug from safe code.
Writing such checkers by hand requires expertise in both the target codebase and the analysis framework, making manual authoring expensive and difficult to scale.

LLM-based patch-driven synthesis offers a promising alternative.
In this paradigm, a security patch serves as a compact before-and-after specification: the LLM consumes the patch and synthesizes a checker that should fire on the vulnerable revision but remain silent on the fixed revision.
KNighter~\cite{yang2025knighter}, a representative system, has demonstrated that this approach can produce compilable and executable CSA checkers from historical Linux kernel fixes.

However, a compilable checker is not necessarily a correct one.
The difficulty is that a patch shows what changed in the code, but not why the change makes the code safe.
Consider a one-character fix that tightens a loop bound from \lstinline|i < N| to \lstinline|i < N-1|: the real reason this works is that it prevents an array access at \lstinline|i+1| from exceeding the array's capacity.
A checker that merely looks for the old loop bound but does not understand the array access behind it will either report both versions or detect neither.
Attempting to fix such a checker can also backfire: edits that suppress spurious reports may simultaneously remove the trigger for the true bug, a failure mode we call over-pruning.

A natural response is to iteratively refine the checker using feedback.
Existing strategies, including KNighter's own loop, feed compilation errors and validation reports back to the LLM so it can edit the checker~\cite{yang2025knighter,madaan2023selfrefine,shinn2023reflexion}.
Yet this feedback only describes symptoms. A message such as ``the vulnerable side is silent'' tells the model that something is wrong, but not what the checker misunderstands about the code.
A cleanup attribute may name a lowered callback that the checker never recognizes. A cast may hide the expression expected by a callback. A state lookup may use a different region from the preceding update.

Our key observation is that \textbf{the analysis framework itself already computes the semantic information needed to close this gap, but existing refinement loops do not use it}.
CSA, for example, exposes control-flow graphs, symbolic values, memory regions and checker callbacks. CodeQL provides similar depth through its program database and queries~\cite{avgustinov2016ql,codeqlDocs}.
This information can reveal why a checker misses a vulnerability, not just that it does.
The challenge is selecting the relevant parts of this rich output and relating them to the patch and the checker under refinement.

We present \toolname, a refinement framework that bridges this gap by feeding analyzer-internal evidence into the LLM's editing loop.
Where existing refinement loops rely primarily on compilation and report feedback, \toolname additionally extracts semantic context from the analyzer itself: a rule-guided preflight selects patch-relevant views such as control-flow structure, symbolic-value bindings and checker callback traces, so the model can see what the checker actually observes at runtime rather than guess from silence alone.
These facts are assembled with source context and validation feedback into a typed evidence bundle with explicit provenance, giving the model a structured diagnosis instead of a flat error log.
The agent then edits the checker and validates it on both revisions, guarded by a retention gate that accepts warning recovery or false-positive reduction but rejects edits that lose an existing vulnerable-side warning, directly countering the over-pruning problem.

This paper makes the following contributions:
\begin{itemize}[leftmargin=*]
  \item We identify a gap between paired warning feedback and the analyzer's representation of the code a generated checker inspects. We define analyzer-internal evidence by its backend origin, extraction interface and source binding, and study how it can guide edits to an existing checker rather than regenerate one.
  \item To address this gap, we build \toolname as an end-to-end plug-in refinement system. It selects patch-relevant views, joins analyzer, source and validation records in a typed, provenance-labelled bundle, and iteratively edits, compiles and scans the detector on both revisions. A retention gate accepts warning recovery or fixed-side report reduction without losing an existing vulnerable warning. The main implementation targets CSA, with a separate CodeQL query adapter.
  \item We then evaluate all 39 selected generated-only Linux CSA checkers from the same starting artifacts, comparing the actual KNighter refinement loop with native and no-internal refinement. Native recovers eight initially silent vulnerable revisions; the ablation without internal evidence recovers six, with each arm repairing cases the other does not, indicating that internal and source-derived evidence address complementary recognition gaps. We also repeat the evidence comparison on 12 subjects, run three editor configurations three times on that subset, refine an archived CodeQL query in three decodes, and inspect all 11 non-tied comparisons alongside controlled source changes. These analyses trace each warning-count change to a specific checker repair, such as recognizing a lowered callback or unwrapping an expression cast.
\end{itemize}

%% file: sections/02-background.tex
This section introduces the background on static vulnerability checkers, LLM-based checker generation from security patches, and the analyzer-internal intermediate results that motivate our approach.

\subsection{Static Vulnerability Checkers}

Static analysis examines source code without executing it, aiming to find bugs, security vulnerabilities, and code-quality issues at compile time or during code review~\cite{engler2001bugs,bessey2010coverity}.
A vulnerability checker is a rule or query within a static-analysis framework that flags program locations matching a security-relevant condition.
Modern frameworks support different checker paradigms.
Path-sensitive engines such as CSA explore program paths, tracking branch guards, symbolic values and memory regions~\cite{clangsaDocs}.
Relational query engines such as CodeQL express checkers as queries over extracted program facts~\cite{avgustinov2016ql,codeqlDocs}.
Both paradigms have been deployed at industrial scale~\cite{sadowski2015tricorder,calcagno2015infer}.

Even so, writing effective checkers remains a bottleneck.
A checker must encode the conditions that distinguish a real bug from a safe use of the same operation.
It must also express those conditions through the framework's callbacks, values and object representations.
The resulting warning burden affects practical use~\cite{johnson2013why,heckman2011alerts}.
Warning ranking and actionability prediction help users inspect reports~\cite{kremenek2002zranking,ruthruff2008actionable}; checker refinement changes the code that produces them.

\subsection{LLM-Based Patch-Driven Checker Generation}

A security patch provides a compact before-and-after specification: the vulnerable revision should trigger the generated checker, and the fixed revision should not.
This idea builds on work that treats patches as vulnerability specifications, including semantic patching~\cite{padioleau2008coccinelle}, clone-based vulnerability discovery~\cite{jang2012redebug,kim2017vuddy}, and patch-derived datasets~\cite{fan2020bigvul}.
KNighter~\cite{yang2025knighter} synthesizes CSA checkers from historical Linux fixes and already refines them using compiler feedback, diagnostic and report-path context, and validation on both revisions.
Its no-report branch provides no report-driven edit.

A patch describes a particular repair rather than every safe and unsafe use of an operation.
The generated checker may therefore omit an important condition or encode it incorrectly.
\toolname retains the patch-driven setting and adds analyzer context to the editing loop, including recovery attempts when the starting checker is silent.

\subsection{Analyzer-Internal Intermediate Results}

CSA maintains symbolic states and region bindings during path exploration; Clang also exposes control-flow and call-graph representations.
CodeQL materializes program relations that can be queried for call and data-flow information~\cite{clangsaDocs,avgustinov2016ql,codeqlDocs}.
Such representations have long supported taint analysis, graph-based detection and interprocedural reasoning~\cite{livshits2005finding,jovanovic2006pixy,arzt2014flowdroid,yamaguchi2014cpg}.

These facts can help explain how a generated checker interacts with its backend.
For example, a callback observation can reveal the expression shape or region on which the checker actually operates.
Their interpretation depends on their source: a compiler-produced record, an emitted diagnostic and a source window supply different kinds of information.
Section~\ref{sec:evidence-roles} defines these origins and the interfaces used in our implementation.

%% file: sections/03-motivating-example.tex
We illustrate the refinement process end to end on a single case: the KNighter-derived CSA checker for Linux commit \texttt{97cba232549b}, which repairs an off-by-one access in the AMD display driver~\cite{linuxCommit97cba232,yang2025knighter}.
This case is small enough to inspect manually yet exhibits the core problem: the generated checker recognizes a look-ahead access but misses the relation between the loop bound and the array's capacity.

\begin{figure}[!htbp]
  \centering
  \includegraphics[width=0.70\textwidth]{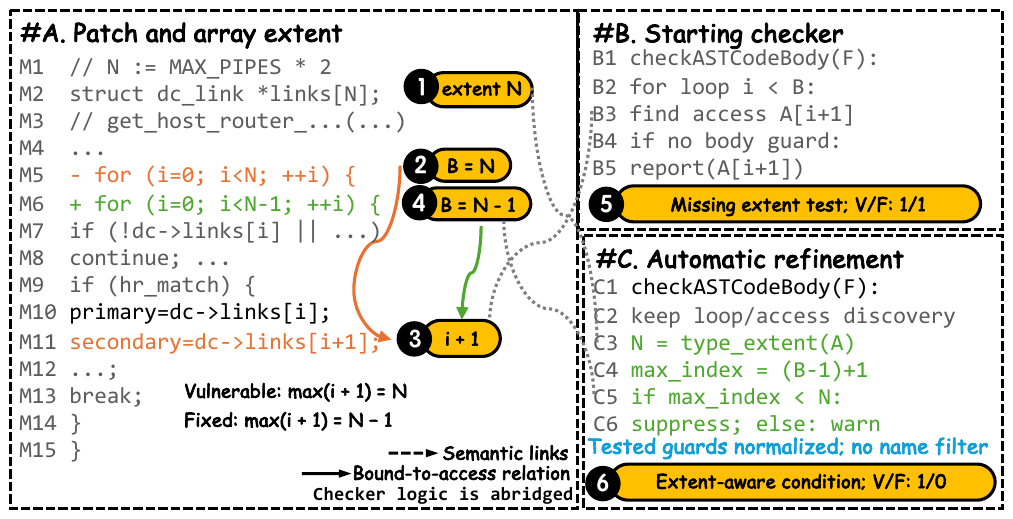}
  \Description{A three-panel example. The left panel shows the array extent
  and the loop-bound patch. The upper-right panel shows the original
  look-ahead checker, which ignores the bound-to-capacity relation. The
  lower-right panel adds the maximum-index comparison while retaining the
  AST-based checker. Solid arrows relate loop bounds to the access; dashed arrows
  connect checker conditions to code.}
  \caption{\textbf{Motivating example.} An off-by-one look-ahead bug and its refinement. $N$ = \lstinline|MAX_PIPES * 2|. Alerts change from \texttt{1/1} to \texttt{1/0} after \toolname adds a capacity-aware condition.}
  \label{fig:motivating}
\end{figure}

\subhead{The bug and its patch.}
The array \lstinline|dc->links| has a declared capacity of \lstinline|MAX_PIPES * 2|, abbreviated as $N$ in Figure~\ref{fig:motivating} (\cmark{1}).
The function \lstinline|get_host_router_total_dp_tunnel_bw| iterates over these links.
On a matching host-router branch, the loop body reads both \lstinline|links[i]| and \lstinline|links[i + 1]| (M10--M11, \cmark{3}).
Under the original bound $i<N$ (M5, \cmark{2}), the loop allows $i=N-1$, so the look-ahead reaches index $N$ and overflows the array.
The patch tightens the bound to $i<N-1$ (M6, \cmark{4}), capping the largest look-ahead index at $N-1$.

\subhead{Why the baseline checker fails.}
The generated checker already recognizes an induction variable, a loop bound, and a look-ahead subscript.
It warns whenever the access lacks a protective condition in an enclosing \lstinline|if| (B1--B5, \cmark{5}).
This rule checks only the branch guards and overlooks the protection carried by the loop bound itself.
Because the fixed program still contains \lstinline|links[i + 1]|, the checker fires on both revisions despite the tighter bound, yielding \texttt{1/1} alerts in independent replay.
KNighter's own refinement loop does not change this checker: its report-driven policy finds no actionable feedback on this case, so the output remains \texttt{1/1}.
In short, the checker has learned that a look-ahead access needs a guard, but not that the loop condition already provides one.

\subhead{Our refinement with analyzer evidence.}
\toolname starts from the same validation symptom (\texttt{1/1}) but goes further by assembling analyzer-internal context alongside the patch and source.
Specifically, it collects four evidence records: two from Clang~18 internals (a CFG dump via \texttt{debug.DumpCFG} and a call graph via \texttt{debug.DumpCallGraph}) and two from analyzer output.
The CFG exposes the loop structure and branch context that the checker misses. The patch and source records identify the bound change.

With this evidence, the LLM adds a capacity-aware condition to the existing \lstinline|checkASTCodeBody| traversal.
The refined checker reads the declared array extent from the AST type and computes the maximum look-ahead index: for a loop $i<B$, that index is $(B-1)+1=B$.
It suppresses the warning when $B$ is smaller than the array extent $N$ (C3--C6, \cmark{6}).
Under the vulnerable bound $B=N$, the maximum index equals the extent, so the warning fires.
Under the fixed bound $B=N-1$, the maximum index is safely below the extent, so the checker stays silent.
Crucially, this condition is based on type and index relations rather than the patched function name, a specific capacity constant, or adjacency of statements.
The first candidate passes independent Linux replay at \texttt{1/0}.

\subhead{Benefits over the baseline.}
The refined checker offers two concrete advantages over KNighter's refined output.
First, higher precision on the original Linux pair: \toolname's checker correctly reports \texttt{1/0} (warns on the vulnerable revision, stays silent on the fixed one), whereas KNighter's checker remains at \texttt{1/1} because it never learned the bound-to-capacity relation.
Second, better generalization: on 114 controlled C programs built from the same look-ahead pattern (with varied names, capacities, unrelated statements, and loop-bound formulations), \toolname's checker gives the expected result on all 114 executions (58 unsafe warned, 56 safe silent), while the checker produced by KNighter's refinement loop correctly handles only 31 unsafe and 25 safe cases. AddressSanitizer independently confirms the safety labels for 80 of these variations~\cite{serebryany2012asan}.
Both advantages stem from the same root cause: analyzer evidence (CFG and call-graph context) lets the model understand why the loop bound matters for safety, so the refined checker encodes a semantic condition (maximum index vs.\ array extent) rather than matching syntactic patterns.


%% file: sections/03-approach.tex
\toolname is a post-synthesis refinement layer for LLM-generated vulnerability detectors.
Given a draft CSA checker or CodeQL query produced by an upstream generator, the patch that motivated the synthesis, and the corresponding source tree and validation scope, \toolname refines the detector to recover vulnerable-side warnings or reduce fixed-side reports.

\begin{figure}[!htbp]
  \centering
  \includegraphics[width=\textwidth]{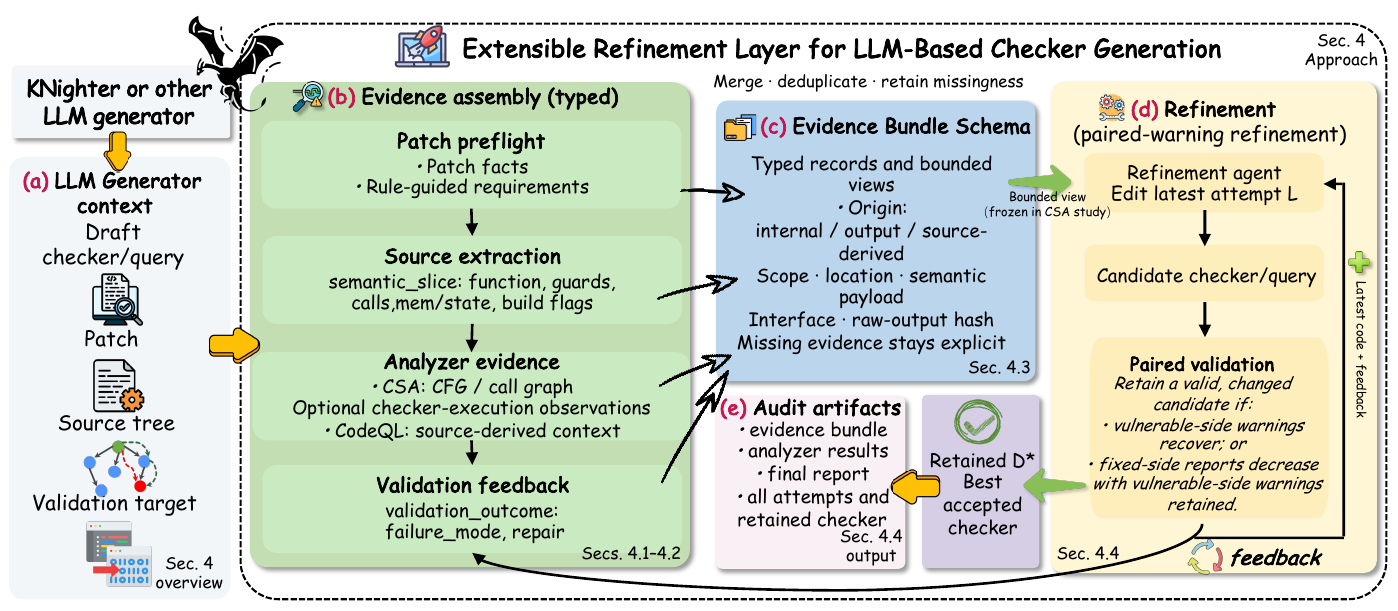}
  \Description{Five labelled components show the upstream checker and patch, evidence assembly, typed bundle, paired-warning editor and audit artifacts. Three evidence collectors and validation feedback feed the bundle. A bounded view goes to the editor; paired validation either updates a separate retained checker or sends the latest code and feedback to the editor. The retained checker points to audit artifacts.}
  \caption{\textbf{\toolname architecture.} Evidence assembly (green) feeds a provenance-labelled bundle (blue) to the editor (yellow). The retention gate (purple) keeps the best checker and returns feedback for the next iteration.}
  \label{fig:approach-architecture}
\end{figure}

Figure~\ref{fig:approach-architecture} summarizes the workflow across five stages.
\textbf{(a)~Generator context.} \toolname receives a draft detector, patch, source tree and validation target. It treats the generator as a black box and refines its output in place.
\textbf{(b)~Evidence assembly.} Rule-guided preflight selects evidence requirements, and four collectors gather patch facts, source context, analyzer evidence and validation feedback.
\textbf{(c)~Evidence bundle schema.} The facts are normalized into typed records with their origin and source bindings.
\textbf{(d)~Paired-Warning Refinement.} The agent edits the latest attempt using a bounded view of those records. Paired validation updates a separate best retained detector.
\textbf{(e)~Audit artifacts.} Each run persists evidence, prompts, replies, candidates and validation results, so a refinement can be inspected or resumed.
The audit links the retained checker to the evidence shown, attempted edits and paired outcomes; RQ3 compares refinement with and without native evidence.

\begin{algorithm}[t]
\footnotesize
\caption{Paired-Warning Evidence-Guided Refinement}
\label{alg:evidence-guided-refinement}
\begin{algorithmic}[1]
\Require detector $D$, patch $P$, source $S$, scope $V$, backends $A$, budget $K$
\Ensure refined detector $D^*$ and audit report
\State $R_0 \gets \Call{Validate}{D,P,V}$
\State $E_p,E_s \gets \Call{Preflight}{P},\ \Call{ExtractSource}{P,S,D}$
\State $E_a \gets \bigcup_{a\in A}\Call{CollectNative}{a,D,P,S}$ \quad $E_v \gets \Call{Feedback}{R_0}$
\State $B \gets \Call{NormalizeMerge}{E_p,E_s,E_a,E_v}$
\State \textbf{if} $\Call{PDS}{D,P,V}$ \textbf{then return} $\Call{Audit}{D,B}$
\State $D^* \gets D$ \quad $L \gets D$ \quad $T \gets \Call{InitialStopState}{}$
\For{$i=1$ to $K$}
  \State $V_i \gets \Call{BoundedView}{B}$ \quad $C_i \gets \Call{AgentEdit}{L,P,V_i}$
  \State $R_i \gets \Call{ValidateIfCompilable}{C_i,P,V}$
  \State \textbf{if} $\Call{Accept}{C_i,D^*}$ \textbf{then} $D^* \gets C_i$ \Comment{Equation~\ref{eq:accept}}
  \State $L \gets C_i$ \quad $E_i \gets \Call{Feedback}{R_i}$
  \State $B \gets \Call{Merge}{B,E_i}$
  \State $T \gets \Call{UpdateStopState}{T,R_i,D^*}$
  \State \textbf{if} $\Call{Stop}{T,K}$ \textbf{then break}
\EndFor
\State \Return $\Call{Audit}{D^*,B}$
\end{algorithmic}
\end{algorithm}

\subsection{Evidence Sources and Roles}
\label{sec:evidence-roles}

\subhead{Analyzer-internal evidence.}
A path-sensitive analyzer does not just emit a final warning: along the way it builds control-flow graphs, tracks symbolic values and region bindings, resolves call edges, and maintains checker-specific state transitions.
These intermediate results are richer than the final diagnostic and cannot be recovered from source code alone.
We call evidence extracted from them \emph{analyzer-internal}, distinguishing it from \emph{source-derived} evidence (patch diffs and source windows) and \emph{analyzer-output} evidence (diagnostics and validation reports).

\subhead{Evidence collection.}
Four collectors run in sequence (Figure~\ref{fig:approach-architecture}b, top to bottom).
\textbf{Patch preflight} parses affected files, hunks and statements, establishing \emph{what changed} and \emph{where}.
\textbf{Source extraction} records bounded contexts around the changed code---enclosing functions, branch guards, callee signatures, memory operations and build flags---providing local semantic context without requiring the full repository.
\textbf{Analyzer evidence} captures intermediate results from the backend.
In our CSA study, the collector invokes Clang~18's \texttt{debug.DumpCFG} and \texttt{debug.DumpCallGraph} to obtain CFG branches, value bindings and call edges, and instruments CSA callbacks to observe abstract values, region bindings and checker-state associations during path exploration.
These are observations of the checker's own analysis, not concrete executions of the program under test.
Instrumented scans must preserve diagnostics on both revisions; we verify revision, source scope and diagnostic parity, and mark unavailable observations explicitly.
\textbf{Validation feedback} records whether the current detector compiles, executes, warns on the vulnerable revision and stays silent on the fixed revision, supplying the failure symptom for the next edit.

\subhead{Provenance classification.}
Each collected record is classified by its origin into three categories (Table~\ref{tab:origin-types}).
\emph{Analyzer-internal} records come from intermediate representations or instrumented callbacks and retain their raw-output hash and source scope.
\emph{Analyzer-output} records contain final diagnostics and paired validation results, with no internal state.
\emph{Source-derived} records hold patch diffs or bounded source windows.
A CodeQL record qualifies as internal only when the database query, revision and returned tuples are retained; the CodeQL subjects in this study use source-derived context.
Across 39 subjects, the initial collection yields 193 records: 81 internal (45 guard, 27 state-transition, 9 allocation-lifecycle), 51 output and 61 source-derived.
Per-attempt availability is reported in RQ3.

\begin{table}[!htbp]
\centering
\caption{Evidence primitives and provenance in the audited 39-case CSA cohort. One record may contain several CFG primitives; counts are by record origin, not primitive.}
\label{tab:origin-types}
\begingroup\FTTableSetup
\begin{tabular}{>{\raggedright\arraybackslash}p{0.27\columnwidth}|>{\raggedright\arraybackslash}p{0.53\columnwidth}|c}
\hline\hline
Origin & What the persisted record contains & Records \\
\hline
Analyzer-internal & Clang CFG guards, value bindings and call edges; raw-output and scope binding & 81 \\
Analyzer output & CSA diagnostics and paired validation reports; no internal state & 51 \\
Source-derived & Patch-diff records here; other interfaces may emit source-window fallbacks & 61 \\
\hline\hline
\end{tabular}
\endgroup
\end{table}

\subhead{Record types.}
The seven persisted types---\texttt{patch\_fact}, \texttt{semantic\_slice}, \texttt{dataflow\_candidate}, \texttt{call\_chain}, \texttt{path\_guard}, \texttt{allocation\_lifecycle} and \texttt{state\_transition}---organize changed statements, bounded source context, possible propagation, call context, control conditions, resources and abstract state, respectively.
When a selected role yields no evidence, the bundle retains an explicit missing-evidence marker so the refinement agent knows what was requested but unavailable.

\subsection{Evidence Selection}
\label{sec:adaptive-selection}

The central design choice in evidence assembly~(Figure~\ref{fig:approach-architecture}b) is \emph{what to collect}.
Analyzer outputs can be large and heterogeneous; passing them wholesale to the model dilutes the facts relevant to the checker under refinement.
\toolname limits collection to patch-related source and presents selected evidence roles through bounded views.

\subhead{Rule-guided preflight.}
Within Patch preflight, the PatchFacts Extractor parses affected functions, added guards, risky calls and type widenings.
Evidence Planner uses a primary pattern from analysis metadata when available, or infers it from these facts.
It assigns fixed priorities: patch anchoring is 100; lifetime and state requirements are 92 for double-free and 89 for use-after-free; null and bounds guards are 90.
Requirements are sorted by decreasing priority, with evidence type breaking ties; the default retains at most eight.
Table~\ref{tab:evidence-routing} shows representative mappings.
These priorities order collection requirements, rather than express confidence in individual facts.
This component uses rules rather than a separately trained or prompted patch-mechanism classifier.

For the CSA comparison, collection is precomputed and the visible view is frozen before editing.
The selected v43 implementation resolves compiler references into at most four function-local CFG witness cards, scoped by the patch and reports.
It compares state associations and helper spelling with the checker, then appends available checker-execution observations.
Additional evidence requests are disabled in this experiment.
The general interface supports bounded, deduplicated role requests; their utility is outside this comparison.

\begin{table}[!htbp]
\centering
\caption{Representative evidence requirements by patch mechanism, using the persisted record types.}
\label{tab:evidence-routing}
\begingroup\FTTableSetup
\begin{tabular}{>{\raggedright\arraybackslash}p{0.30\columnwidth}|>{\raggedright\arraybackslash}p{0.42\columnwidth}|>{\raggedright\arraybackslash}p{0.18\columnwidth}}
\hline\hline
Patch mechanism & Possible record types & Backend \\
\hline
Cleanup edge or double free & \texttt{allocation\_lifecycle}, \texttt{state\_transition}, \texttt{path\_guard} & CSA \\
Bounds widening or overflow & \texttt{semantic\_slice}, \texttt{dataflow\_candidate}, \texttt{state\_transition} & CSA and CodeQL \\
Initialization or null dereference & \texttt{state\_transition}, \texttt{path\_guard} & CSA \\
Use-after-free or escape & \texttt{allocation\_lifecycle}, \texttt{call\_chain} & CSA and CodeQL \\
API misuse or missing check & \texttt{semantic\_slice}, \texttt{call\_chain} & CodeQL \\
\hline\hline
\end{tabular}
\endgroup
\end{table}

\subsection{Evidence Bundle Schema}
\label{sec:evidence-schema}

The normalizer merges all collected facts into the EvidenceBundle shown in Figure~\ref{fig:approach-architecture}c.
Each fact becomes an EvidenceRecord carrying a type, the contributing analyzer, scope, source location, payload, provenance, and an optional EvidenceSlice (the compact semantic content consumed by the agent).
The schema separates program facts from analysis rules in the spirit of relational analyzer frameworks~\cite{avgustinov2016ql,jordan2016souffle,bravenboer2009doop}, but uses a smaller role vocabulary tailored to checker refinement.
Formally:
\begin{equation}
\begin{aligned}
B &= \mathrm{Norm}(E_p \cup E_s \cup E_a \cup E_v),\\
E_a &= E_{\mathrm{CSA}} \cup E_{\mathrm{CodeQL}},\\
r &\in B.records = \langle type, analyzer, scope,\\
&\quad location, payload, provenance, slice\rangle.
\end{aligned}
\label{eq:evidence-bundle}
\end{equation}
Here $E_p$ contains patch facts, $E_s$ bounded source slices, $E_a$ analyzer representations or outputs, and $E_v$ validation outcomes.
The normalizer removes duplicates, records contributing analyzers and retains missing-evidence markers.
Provenance identifies each record as \texttt{analyzer\_\allowbreak internal}, \texttt{analyzer\_\allowbreak output} or \texttt{source\_\allowbreak derived}.
The evidence type and origin are separate: the same role may hold a backend-produced record or a source-derived fallback.
This decouples the bundle from any single backend representation.

\subsection{Paired-Warning Refinement}
\label{sec:refinement}

The refinement agent receives the current detector, the patch, baseline validation notes and a bounded evidence view.
The tool-mediated loop draws on LLM agent and self-refinement architectures, with actions restricted to evidence requests, detector edits and validation commands~\cite{madaan2023selfrefine,shinn2023reflexion,yao2023react,schick2023toolformer}.
The CSA comparison makes zero additional evidence requests.

\subhead{Model and programmatic steps.}
The model proposes checker edits and may supply patch-analysis metadata or enabled role requests.
Preflight rules, normalization, raw-output hashing, source binding, compilation, paired replay and acceptance are programmatic.
The editor prompt gives the current and starting checkers, patch, source and evidence context, symptoms on both revisions, and backend-specific editing and validation rules.
Repair prompts contain the actual compiler diagnostics.
The artifact retains English templates and rendered exchanges, with missing historical exchanges marked.

\subhead{Retention gate.}
A changed candidate must compile and execute normally on both revisions.
Equation~\ref{eq:accept} then retains a recovered vulnerable-side warning, or fewer fixed-side reports while preserving vulnerable-side warning presence.
Fixed-side silence with a vulnerable warning achieves the stronger PDS outcome.
Native refinement additionally checks required hits on controlled array fixtures; Section~\ref{sec:evaluation} describes this guard and the warning-level metrics.

\subhead{Feedback loop.}
When a candidate fails to compile or improve the retained result, its code and feedback remain available for the next attempt.
The agent edits the latest attempt, while $D^*$ stores the best retained checker.
Algorithm~\ref{alg:evidence-guided-refinement} gives the loop; the protocol specifies its stopping conditions and treatment of interrupted executions.
CSA checkers are compiled and replayed against both revisions, while CodeQL queries run over the corresponding databases.

\subhead{Adding a backend.}
An adapter provides registration, evidence and provenance mapping, dependencies, execution and result decoding, and paired validation.
Shared modules still contain backend branches.
The four CodeQL-specific modules occupy 2,438 physical lines, including comments and generation code.
With an analyzer and patch pair already available, we estimate that a familiar developer using Codex could complete a basic adapter within an afternoon; developer time was not measured.
New native-evidence semantics and broad validation require additional work.
A new bug mechanism needs a role mapping and fixtures for safe and unsafe cases, with a new extractor where a required relation is unavailable.

%% file: sections/04-evaluation.tex
We evaluate whether analyzer evidence helps \toolname recover vulnerable-side warnings and reduce fixed-side reports.
The evaluation focuses on patch-local and patch-related-file behavior rather than whole-kernel bug discovery.

\subsection{Research Questions}

\noindent\textbf{RQ1: KNighter-derived checkers.} How does \toolname compare with KNighter's refinement loop on vulnerable-version warning coverage, fixed-side report burden and paired discrimination?

\noindent\textbf{RQ2: Cross-backend feasibility and cost.} Can the refinement layer operate on an archived CodeQL query associated with an earlier automatic generation run, and what integration and execution costs are observable?

\noindent\textbf{RQ3: Evidence contribution.} How do the native and no-internal variants differ, and what do their edits and reported gains establish?

\noindent\textbf{RQ4: Model sensitivity.} How does the workflow perform with three editor configurations on the same subset under common reply and output ceilings?

\subsection{Experimental Setup}

\subhead{Subjects and benchmarks.}
We organize the evaluation into four experiments (E1--E4).
E1 uses KNighter-derived Linux CSA checkers under patch-related object validation.
From 286 artifact records on 60 commits, 241 records on 39 commits lack an upstream refinement-status row.
We select one nonempty repaired-source artifact per commit by upstream TP, then TN, then the lowest checker index.
This fixed, non-random rule selects the inputs; independent paired replay supplies their starting outcomes.
All 39 enter KNighter refinement and both \toolname variants, including cases that remain noisy or fail to improve.
Initially, 15 checkers warn on both revisions and 24 are silent on both.
Their zero starting PDS is the study's entry state.

The 15 initially noisy checkers also form a complete report-refinement stratum.
Its results reuse the full-cohort outputs.
For the 24 silent starts, KNighter's no-report branch compiles without a model edit, whereas \toolname attempts recovery.
We report these branches as the systems execute them.

The selected implementation is v43, chosen after comparing two configurations on six experimental subjects.
Compatible executions on those subjects are reused in E1; the configuration-selection cases overlap the evaluation.
The artifact retains those trials and their reuse checks.
G04 uses its architecture-correct ARM64 scope and evidence after repair of an earlier wrong-object scan.

\begin{figure}[!htbp]
\centering
\includegraphics[width=0.70\textwidth]{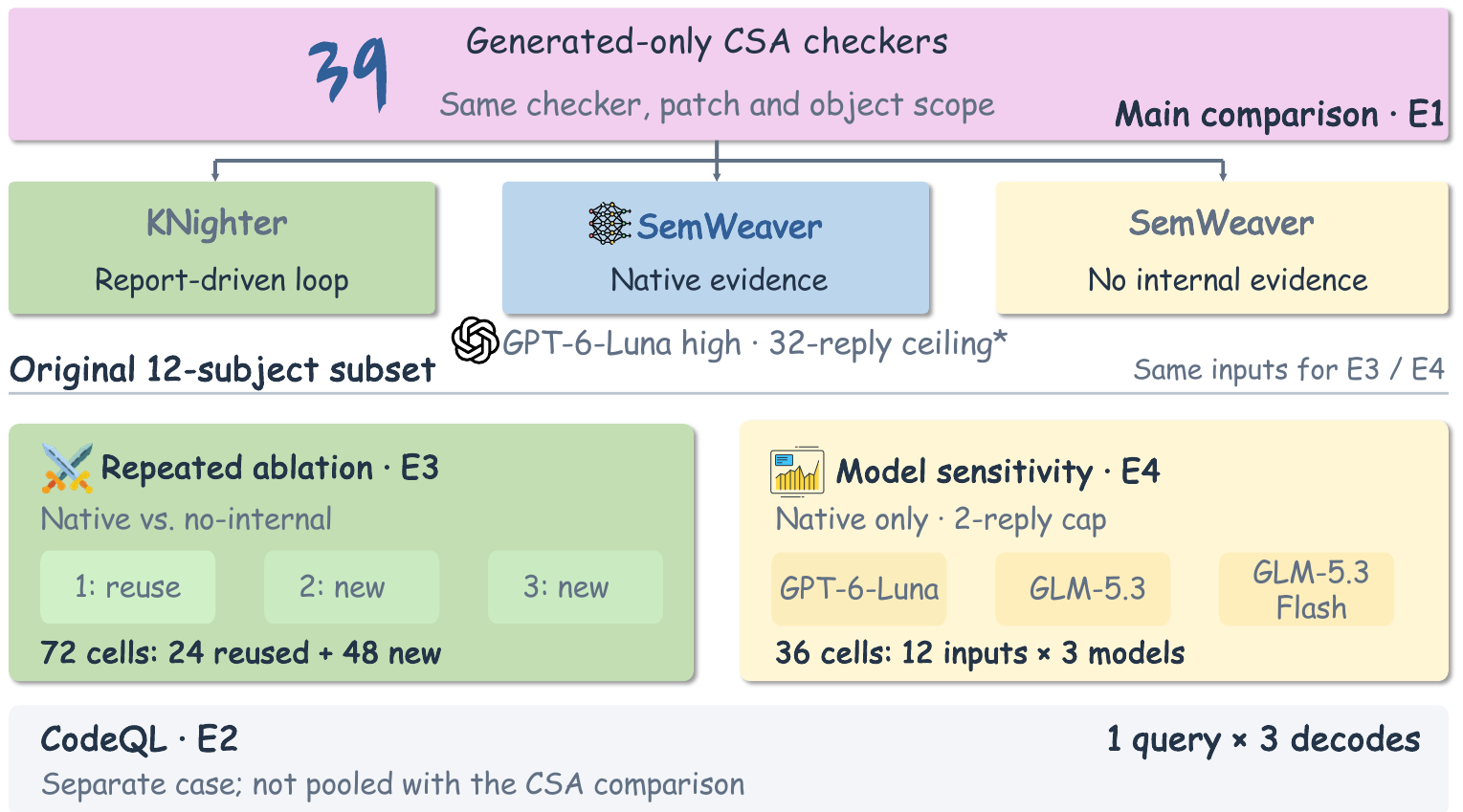}
\Description{All 39 CSA checkers enter three main comparison arms. E3 and E4 use the same original 12-subject subset. E3 reuses 24 first-decode condition cells and adds 48. The drawing shows E4's first 36 native-only cells; two further decodes add 72, for 108 trial records. The CodeQL case is separate.}
\caption{\textbf{Experimental design and reuse.} E3's first decode reuses E1. The E4 panel depicts its first decode (36 cells); two further decodes add 72 cells, for 108 records on the same 12 inputs. Cells are not independent bugs. $^*$Audited natural-stop reuses retain original caps; shared ceilings do not imply equal realized effort.}
\label{fig:study-design}
\end{figure}

E2 uses an archived CodeQL input associated with an earlier automatic generation run, independently of KNighter, for ImageMagick (CVE-2017-12877, \texttt{coders/\allowbreak mat.c}).
All three refinement decodes start from this same query.
We report it separately from CSA.
Earlier 20-row results from mixed and manual inputs are excluded.
E3 compares native and no-internal on all 39 main pairs and repeats a 12-subject subset; E4 uses the same 12 subjects, native only, with three decodes per model (Figure~\ref{fig:study-design}).

Subset membership follows the initial scan status.
We sort SHA-256 hashes of a fixed artifact-recorded salt plus case ID within the original 14 noisy and 25 silent cases, then take the first 4 and 8.
A later scan repair corrects G08 from $0/0$ to $2/2$, leaving the same membership with 5 noisy and 7 silent subjects.
The selection uses no refinement outcomes or vulnerability-class stratification; it contains no bounds cases.

\subhead{Baselines and variants.}
We hold the generated detector fixed so that the comparison follows edits to the same checker rather than a fresh generation draw.
E1 native, no-internal and KNighter use the same requested GPT-6-Luna-high editor and a 32-reply ceiling.
KNighter retains its report-driven algorithm with five reports and two outer attempts, and its emitted checkers receive independent scans.
The 24 no-report branches consume no reply.
Two natural one-response No-FP stops are audited reuses that retain their original eight-reply ceiling; both terminate before either ceiling.
We report actual replies and warning counts alongside the common maximum.

Within \toolname, no-internal removes native compiler and checker-execution views while keeping the checker, patch, frozen source, build and paired-scan feedback, and stopping rule.
Native receives the bounded CFG witness and mismatch views and available dynamic observations.
E3 repeats the original 12 subjects at 32 replies and 16,384 output tokens per call.
E4 uses GPT-6-Luna high, GLM-5.3 high and GLM-5.3-Flash high, with three decodes per model, 2 replies per decode and 65,536 output tokens per call.
Actual GLM effort fields are retained.
Earlier default-effort, smaller-output and outside-subset histories are kept separate.
E4 has no no-internal arm across models.

\subhead{Experimental procedure.}
For each sample, we collect evidence, refine the detector and validate it on both revisions.
Each permitted reply counts as an attempt.
Compilable changed candidates receive independent paired replay and are retained under Equation~\ref{eq:accept}.
The next call receives the latest attempted code and feedback; the best retained detector is stored separately.
The loop stops when a vulnerable warning and fixed-side silence are achieved, after three valid validations without improvement, after three consecutive invalid model attempts, or at the reply cap.

We preserve failed attempts and consumed prefixes when repairing execution problems.
Thinking-only replies consume budget.
For one GLM reply containing an explicit action and three edits but no explanatory summary, network-disabled replay supplied an empty summary and reproduced the model-specified code with no new call.
Healthy zero-callback probes use static evidence.
Compilation failures, format failures, unavailable observations, provider interruptions and invalid scans remain distinct; no manual checker completion enters the scored results.

\subhead{Environment.}
CSA uses Clang/LLVM 18.1.3 and Python 3.12.3.
CodeQL uses version 2.26.4, \texttt{codeql/cpp-all} 8.0.3 and no-build databases.
Each Linux subject uses its patch commit and parent, \texttt{allyesconfig}, architecture-correct objects, and the same scope for starting, baseline and refined checkers.
Manifests record code, evidence and raw paired outputs.
The recorded model labels and high-effort requests identify the tested services.
Both Responses adapters omit the configured temperature field, and no fixed seed is used.

\subhead{Use of generative AI in the study.}
The editor models generated the checker and query edits evaluated in this study.
Codex assisted with the development of experimental tools.

\subsection{Metrics}

We report vulnerable-version warning presence and fixed-side report burden.
Let $n_v(D)$ and $n_f(D)$ count warnings in the same patch-related object scope on the vulnerable and fixed revisions.
The stronger outcome, \emph{patch-discriminating success} (PDS), requires a vulnerable-side warning and fixed-side silence:
\begin{equation}
\begin{aligned}
H_v(D,P,V)&=\mathbb{1}[\mathrm{alerts}(D,V_{\mathrm{vuln}},P)>0],\\
S_f(D,P,V)&=\mathbb{1}[\mathrm{alerts}(D,V_{\mathrm{fix}},P)=0],\\
\mathrm{PDS}(D,P,V)&=H_v(D,P,V)\land S_f(D,P,V).
\end{aligned}
\label{eq:pds}
\end{equation}
We distinguish two useful improvements.
\emph{Warning recovery} makes an initially silent vulnerable revision warning-positive, possibly with fixed-side noise.
\emph{Report reduction} lowers fixed-side reports while preserving vulnerable-side warning presence.
A detector that loses an existing vulnerable warning is rejected.
A comparative disadvantage means that another arm retains a better result, which may still be an improvement over the starting checker.

For retained detector $D^*$ and normally paired-executed candidate $C$, the improvement rule is:
\begin{equation}
\begin{aligned}
\mathrm{Accept}(C,D^*)={}&\mathrm{AnalyzerOK}(C)\land H_v(C)\\
&\land\bigl(\neg H_v(D^*)\lor n_f(C)<n_f(D^*)\bigr)\land Q(C).
\end{aligned}
\label{eq:accept}
\end{equation}
The editor requires a changed artifact before paired assessment.
$Q$ preserves required hits on controlled array fixtures in native and is identically true for no-internal.
This native-only guard records zero rejections in E1/E3.

For precision, recall and $F_1=2TP/(2TP+FP+FN)$, each vulnerable revision is a positive and its fixed revision a negative, classified by warning presence.
These metrics cover 78 patch-version decisions.
Fixed-side report counts measure inspection burden.
Neither measure adjudicates individual bug sites: G08, for example, warns on allocation failure rather than the registration-failure leak repaired by its patch.
The qualitative follow-up in Section~\ref{sec:target-results} examines target association separately; failed scans remain unscored.

%% file: sections/05-results.tex
\subsection{RQ1: KNighter-Derived Checkers}

Table~\ref{tab:all39-system} reports the complete 39-subject comparison.
All three refinement portfolios retain the 15 initially warning-positive vulnerable revisions.
Among the 24 silent starts, native \toolname recovers eight, no-internal recovers six, and KNighter makes no recovery edit under its no-report policy.
We record every subject and its paired counts.

\begin{table}[!htbp]
\centering
\caption{CSA comparison: (a) all 39 subjects; (b) the complete preexisting 15-case fixed-noisy stratum, reusing the same outputs. TP, FP and FN are version-warning decisions; fixed reports measure burden and replies count received responses. Color shades $1-F_1$ at .25, .50 and .75, not significance.}
\label{tab:all39-system}
\label{tab:matched14}
\begingroup\FTTableSetup
\begin{tabular}{l|cccc|c|cc}
\multicolumn{8}{l}{(a) Full cohort: 39 subjects}\\
\hline\hline
\multirow{2}{*}{Condition}&\multicolumn{4}{c|}{Version outcomes}&
\multirow{2}{*}{$F_1$}&\multicolumn{2}{c}{Burden and effort}\\
\cline{2-5}\cline{7-8}
&PDS&TP&FP&FN&&Fixed reports&Replies\\\hline
Unmodified &0&15&15&24&\FTFScore{.435}&73&0\\\hline
KNighter loop &2&15&13&24&\FTFScore{.448}&69&179\\\hline
\toolname, native &5&23&18&16&\FTFScore{.5750}&123&150\\\hline
\toolname, no internal &8&21&13&18&\FTFScore{.5753}&68&133\\
\hline\hline
\end{tabular}
\par\smallskip
\begin{tabular}{l|c|c|c}
\multicolumn{4}{l}{(b) Noisy 15: all methods retain TP=15, PDS=2 and $F_1=.6977$}\\
\hline\hline
& KNighter loop & Native & No internal\\\hline
Fixed reports &69&44&68\\
\hline\hline
\end{tabular}
\endgroup
\end{table}

However, the additional warning coverage comes with more fixed-side reports.
Native produces 123, compared with 68 for no-internal and 69 for KNighter.
Its precision is $0.561$, versus $0.618$ for no-internal, reflecting a trade-off between broader recovery and tighter fixed-side filtering.
G12 accounts for 73 native fixed-side reports at distinct locations.
Its broader matching produces 74 vulnerable reports and 73 fixed reports, showing why warning recovery and report reduction need to be read together.

The initially noisy group gives a different view (Table~\ref{tab:matched14}b).
On all 15 subjects, native reduces fixed reports by 25 (36.2\%) against KNighter and 24 (35.3\%) against no-internal.
The reduction is concentrated in G24's $27\rightarrow2$ fixed reports; other small differences offset one another.
All methods have the same binary fixed positivity and $F_1$ in this group.
Section~\ref{sec:target-results} examines whether G24's surviving reports preserve the intended lifecycle target.

\subsection{RQ2: Cross-Backend Feasibility}

Table~\ref{tab:codeql-decoding} reports three automatic decodes of the archived ImageMagick CodeQL query.
The starting query returns $0/0$ rows on the vulnerable and fixed revisions.
Two decodes recover $8/0$; the third returns $22/22$.
The successful edit corrects relative-path matching while keeping the exact file, function and local-variable filters.
A second query already starts at $2/0$ and is excluded from the gain count.
This experiment exercises query refinement separately from the KNighter CSA comparison.

\begin{table}[!htbp]
\centering
\caption{Three automatic CodeQL decodes of one query. Counts are output rows on the vulnerable and fixed revisions; time is recorded agent-round duration, excluding separate paired replay.}
\label{tab:codeql-decoding}
\begingroup\FTTableSetup
\begin{tabular}{c|c|crr}
\hline\hline
Decode & Rows (V,F) & Replies & Tokens & Agent time (s)\\\hline
1 &8/0&1&15,122&87.87\\
2 &8/0&1&14,345&42.74\\
3 &22/22&8&110,282&516.77\\
\hline\hline
\end{tabular}\endgroup
\end{table}

\subhead{Second-backend costs.}
Three warm paired replays per query reproduce $0/0$ before refinement and $8/0$ after it.
Median query-plus-decode time is 12.57\,s initially and 36.51\,s after refinement with two threads; peak process RSS is 976/1,078\,MiB.
Database-creation log intervals span 31/30\,s, excluding unlogged startup.
These are execution measurements in the existing environment.
Developer labor and cold installation were not measured.

\subsection{RQ3: Contribution of Analyzer Evidence}

Table~\ref{tab:origin-types} accounts for the initial records.
During refinement, 103 of 150 unique native-chain attempts have usable hash-bound checker observations.
The remaining attempts are unavailable because of candidate compilation (11), unsupported context (18), zero scoped callbacks (13), a reproduced candidate crash (1), or capped optional observations (4).
Zero observations are treated as absence of evidence.

Turning to the refinement outcomes, native has six comparative benefits, five disadvantages and 28 ties or other outcomes across the 39 subjects.
The edits include cleanup-wrapper recognition in G19, nested dereference and binding recovery in G25, and handling of null values, regions and frames in G24.
For example, adding the lowered cleanup name \texttt{\_\_free\_kfree} in G19 changes the counts on the vulnerable and fixed revisions to $15/0$, while no-internal remains at $0/0$.
Conversely, no-internal recovers symbolic-region handling in G23 and persistent-field release recognition in G38.
In G12, removing CPU-role restrictions broadens matching and produces the noise discussed above.

\begin{figure}[!htbp]
\centering
\includegraphics[width=378bp]{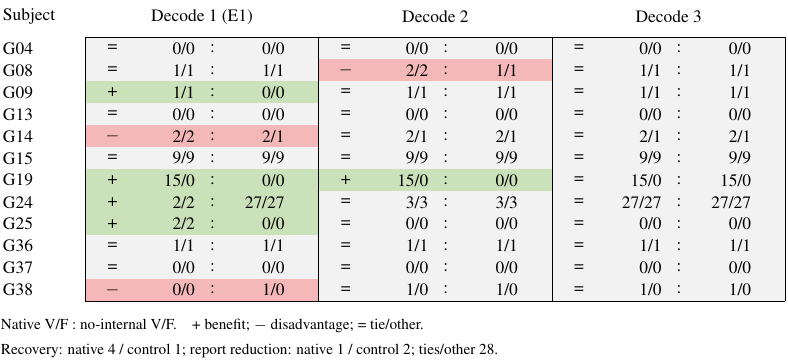}
\Description{A twelve-subject by three-decode matrix of native and no-internal warning counts on vulnerable and fixed revisions, highlighting comparative benefits, disadvantages and ties.}
\caption{E3: all 36 paired decodes on 12 subjects; cells compare native and no-internal counts on the vulnerable and fixed revisions. Decode 1 reuses E1; the remaining decodes add 48 condition cells. Color and sign encode the comparative direction.}
\label{fig:repeats}
\end{figure}

To examine variation across decodes, Figure~\ref{fig:repeats} shows all three paired decodes on the original 12 subjects.
Five subjects tie in every decode; seven show mixed directions across decodes, consistent with the stochastic nature of LLM editing.
G19's native $15/0$ repeats, but no-internal catches up in the third decode.
G24's main report reduction does not repeat.
G14 improves its own start from $3/3$ to $2/2$ but trails no-internal's $2/1$; later decodes tie.
The disadvantages in G05/G23/G38 arise because native stays at its initial $0/0$ while no-internal recovers warnings.
These outcomes show variation in what the editor can recover, even when the accepted artifact preserves its own initial warning signal.

\subsection{RQ4: Sensitivity Across Models}

Table~\ref{tab:models39} uses the same 12 subjects with three native-editor configurations and three decodes each.
All configurations retain the five initial vulnerable warning signals and recover additional versions in the first decode.
The other decodes show how much those results vary.

\begin{table}[!htbp]
\centering
\caption{Limited-budget native sensitivity: three decodes on the same 12 subjects per model, not full-39 robustness. $F_1$ is scored per decode; mean and sample SD summarize the three values. Fixed reports count warning occurrences on fixed revisions.}
\label{tab:models39}
\begingroup\FTTableSetup
\begin{tabular}{l|ccc|cc|c}
\hline\hline
\rule{0pt}{10.2bp}Editor & $F_1$ 1 & $F_1$ 2 & $F_1$ 3 & Mean & SD & Fixed reports 1/2/3\\
\hline
GPT-6-Luna high &.560&.615&.500&.558&.058&40/36/41\\\hline
GLM-5.3 high &.583&.615&.667&.622&.042&16/13/14\\\hline
GLM-5.3-Flash high &.692&.560&.667&.640&.070&11/41/40\\
\hline\hline
\end{tabular}
\endgroup
\end{table}

Received replies across the three decodes are 24/23/24 for GPT-6-Luna, 23/23/21 for GLM-5.3 and 21/23/21 for GLM-5.3-Flash.
Their PDS counts are 1/2/0, 2/2/4 and 4/1/4, respectively; identical warning counts on the vulnerable and fixed revisions across all three runs occur on 8, 3 and 4 of the 12 subjects.
Capped, format-invalid and compile-failed attempts remain charged; a candidate that cannot complete a valid paired scan is not scored as an improvement.
GLM-5.3-Flash has the highest mean $F_1$. However, its first-decode fixed-report advantage does not recur; GLM-5.3 has fewer fixed-side reports in all three decodes.
These results confirm that the refinement framework generalizes across editor models, with the cross-decode variation motivating repeated evaluation in future larger-scale studies.

\subhead{Controlled source replay.}
Beyond model variation, we examine source changes in three selected original subjects, G01, G08 and G09; their unchanged-source controls reproduce the original paired counts.
Local renaming and inert no-ops leave counts unchanged in these cases.
Under the equivalent guard rewrite, native retains vulnerable-side warning presence in eight of nine case--change pairs: G01 falls from $2/2$ to $0/0$, G08 rises from $1/1$ to $2/2$, and G09 remains $1/1$.
Thus the loss of two fixed-side reports in G01 is also a loss of its vulnerable warning, while G08's preserved warning costs another fixed-side report.
These controlled changes identify source-form sensitivity without treating the three subjects as independent natural evolution events.

\subsection{Qualitative Findings}
\label{sec:target-results}

To validate whether warning-level improvements reflect genuine semantic understanding, we conduct three supplementary analyses.
First, a target and report audit inspects all 11 non-tied native/no-internal comparisons across 362 report occurrences and 22 code diffs, classifying each report by target association and source support.
Second, mechanism probes construct six controlled variants per case for G12, G19, G24 and G25 (48 files, 96 scans), and a separate intervention replaces native-observation lines with missing markers to test whether evidence removal changes the model's editing decisions.
Third, a controlled source replay rescans three subjects under local renames, inert no-ops and equivalent guard rewrites to measure source-form sensitivity without re-invoking the model.

To explain the native and no-internal differences, we inspect all 11 non-tied comparisons and their reported locations.
The quantitative results match several distinct changes in the checkers.
Among the six native benefits, G12/G19/G25 have bounded source-supported target reports, G09 has conditional support, and G24/G35 retain no supported or conditional target report.
Among the five no-internal advantages, G01/G23/G38 have source support and G05/G14 conditional support.
This source audit explains the warning counts without changing the primary scores.

Both G01 variants repair recognition of the governing loop, but only no-internal checks the preceding capacity guard.
G19 adds a lowered cleanup name.
G23 changes symbol and nested-member lookup; G25 changes sink recognition and destination binding.
These repairs make the intended operation visible to the checker.
Other edits relax predicates: G05 preserves initialization uncertainty, G09 broadens cleanup handling, G12 drops actor restrictions, and G38 combines field-region preservation with a looser usage gate.
G35's new trigger instead reports on an initialized overflow parameter outside the patched target.

The surviving report population matters as much as the count.
G14 chooses different event predicates and retains reports about different lifetime properties.
G24 changes region, NULL and frame tests, removing 25 reports per side; its two survivors concern owner destruction outside the target surviving-owner lifecycle.
Some removed target reports are conditional.
This case illustrates the trade-off between reducing inspection burden and preserving target coverage, a distinction that warning counts alone cannot capture.

\subhead{Frozen perturbations.}
On the six positive and six negative fixtures per mechanism, retained native G12 reports at a target location in all positives and no negatives.
This is within its annotation model; the diagnostic's remote-write note points to a local reset.
G19 is silent on its adapted-wrapper suite.
Its separate canonical control produces one supported early-exit report and one wrong-scope allocator-return report, and it misses the renamed wrapper.
G25 reports at target locations in all six positives and five negatives.
The starting and no-internal artifacts are silent for these three cases.
For G24, the starting and no-internal artifacts report at target locations in all six positives and six negatives, with 108/108 total reports.
Native has no frozen target-line hits and 18/12 total reports.
Secondary inspection supports some guard warnings outside that line list; the original location scores remain unchanged.

\subhead{Selected-clue intervention.}
All three full-evidence G25 decisions peel the outer casts around the sink expression; no withheld decision does.
Two full decisions also change bookkeeping.
After the separate compiler-only completion, full recovers target-line presence with negative-side noise, while withheld remains silent (Table~\ref{tab:checkpoint-intervention}).
The full-evidence reports still need inspection.
All 34 reports from repetitions 2/3 follow \texttt{argc<=1}, which selects a concrete non-NULL object.
Repetition 1 has six outer count-access reports that admit a source-feasible NULL completion, without an explicit NULL-return event, alongside unsupported reports.

\begin{table}[!htbp]
\centering
\caption{GLM-5.3-Flash bounded-adapter terminal artifacts after separate compiler-only completion. Each cell gives full and withheld results. Primary suites have six positive and six negative fixtures; G19 lowering has two of each and benign has four negatives. Hits and alerts are final-location presence, not validated diagnoses. Original strict-schema and adapter failures remain separate (Section~\ref{sec:target-results}).}
\label{tab:checkpoint-intervention}
\begingroup\FTTableSetup
\setlength{\extrarowheight}{1bp}
\begin{tabular}{c|ccc|ccc}
\hline\hline
\multirow{3}{*}{Draw}&\multicolumn{3}{c|}{G19}&\multicolumn{3}{c}{G25}\\
\cline{2-7}
&Primary&Lowering&Benign&Positive&Negative&Negative\\
&positive hits&positive hits&alerts&hits&alerts&reports\\\hline
1&0/6&1/2&0/2&6/0&5/0&28/0\\
2&0/6&1/2&0/2&6/0&5/0&10/0\\
3&6/0&2/0&1/0&6/0&6/0&6/0\\
\hline\hline
\end{tabular}\endgroup
\end{table}

G19 exposes a different tradeoff.
Full repetitions 1/2 retain name-family tests, while withheld generalizes registration; repetition 3 reverses primary coverage.
The broader matchers can warn on benign cleanup.
Across these artifacts, supported owning-scope exits coexist with reports at callee returns and a write-only cleanup mistaken for a read.

The negative cases are equally important.
Correcting the callback or expression recognized by a checker can activate a useful check, yet leave the object, guard or owning scope wrong.
Even the final source line can contain both a safe pointer-field read and the optional-object dereference of interest.
Conversely, a source-supported guard warning can fall outside the chosen sink-line list.
Examining the edit and its diagnostic together explains why recovery, target-line agreement and semantic support sometimes disagree.

%% file: sections/06-discussion.tex
\subhead{Scope of improvement.}
\toolname refines a detector with respect to its motivating patch.
A human-written checker could use the same interface with a compatible backend, patch and validation pair.
We study generated inputs, for which the patch supplies a common refinement target.
A checker may know \emph{where} to warn yet miss \emph{why} the warning should disappear after the fix.
Our results show that refinement also has an earlier task: making the checker recognize the relevant operation in the analyzer's representation.
G19 needs the lowered cleanup callback; G25 needs to look through an expression wrapper.
These are useful repairs, but the remaining scope and path conditions still determine whether a warning is justified.

\subhead{Evidence as context compression.}
The relevant condition may span source code and analyzer outputs.
Passing the full repository to the model is impractical, while passing only the patch may leave out the representation seen by the checker.
\toolname collects a provenance-tagged summary of patch-relevant facts and makes it available with validation feedback.
The interface follows the general idea of refining an artifact under additional program evidence~\cite{clarke2000cegar,solarlezama2006sketch,nguyen2013semfix,mechtaev2016angelix,long2016prophet}.
The native and no-internal arms repair different representation problems (e.g., G19/G25 vs.\ G01/G23/G38), suggesting that analyzer-internal and source-derived evidence address complementary gaps.

\subhead{Backend interfaces.}
CSA and CodeQL offer different checker formats and evidence interfaces.
The automatic CodeQL case shows that the same editing and paired-validation workflow can operate on a query.
It retains source-derived context and inherited patch-local filters.
The recorded warm execution costs and existing adapter footprint describe that implementation; broader transfer and measured developer effort remain open.

\subhead{Failure modes.}
Over-pruning is one failure mode, but retaining a warning is not enough either.
In G24, fewer fixed-side reports coexist with surviving reports outside the intended lifecycle.
In G19, a broader cleanup matcher can report at a callee return while the caller's variable is still live.
In G25, a recovered target-line warning can follow an explicitly non-NULL path.
These cases suggest reporting the actual checker edit, the property supported by its diagnostic, and the change in inspection burden together.
This separates successful recognition from the semantic conditions still needing repair.

\subhead{Generalization.}
The motivating example survives the tested renamings, capacities and equivalent guard forms.
Those tests informed its refinement, and the additional four-mechanism suites are controlled source reductions.
The replay of three original subjects adds a different check: local renames and no-ops preserve their counts, but an equivalent guard rewrite silences G01 on both versions and increases G08's fixed-side reports.
Transfer to unseen projects and vulnerability instances requires a separate evaluation.

%% file: sections/07-threats.tex
\subhead{Internal.}
An error in evidence collection, editing or validation could produce a spurious improvement.
To mitigate this, we retain raw outputs, code hashes, failed attempts and unavailable observations for every run; all scored checker edits are fully automatic and reproducible from the persisted artifacts.
Native context bundles several components, which limits attribution to any single evidence type.
The checkpoint intervention (Section~\ref{sec:target-results}) isolates selected clues while holding other context constant, and the fixture guard present in native but absent from no-internal rejects no E1/E3 candidate.
Decoding variance from model aliases and the absence of a fixed seed is partially addressed by running three independent decodes per configuration; the variation across decodes is reported explicitly in RQ3 and RQ4.

\subhead{Construct.}
The primary metrics describe warnings on vulnerable and fixed revisions under the same patch-related scope, not target-level precision or confirmed false-positive bugs.
This is a deliberate choice: version-level warning presence is the metric available without manual triage of every report, and it aligns with the paired-validation protocol that upstream generators already use.
The qualitative follow-up (Section~\ref{sec:target-results}) adds source-level target audits and mechanism probes to assess whether warning-count changes reflect genuine semantic repairs; these judgments complement the quantitative scores without replacing them.
Final-line matching cannot distinguish two objects or a feasible and infeasible path on the same line; we note this when interpreting individual cases (G08, G24) rather than treating all warnings as equivalent.

\subhead{External.}
The 39 subjects come from one generator's Linux artifacts, which bounds the diversity of checker styles and vulnerability types.
Within this scope, the subjects cover five patch mechanisms (Table~\ref{tab:evidence-routing}), exercise both silent and noisy starting conditions, and include cases where internal evidence helps, hurts, or makes no difference.
The model study uses three decodes per configuration on 12 subjects, sufficient to reveal cross-decode variation but not to rank models with statistical confidence.
The CodeQL experiment demonstrates cross-backend feasibility on one patch-local example; broader transfer remains future work.
The controlled source replay covers three subjects and tests source-form sensitivity rather than representing independent natural evolution.

\subhead{Conclusion.}
Repeated decodes and transformed fixtures are correlated observations, not additional independent bugs; we interpret them as stability checks rather than as expanding the effective sample size.
The findings characterize local repairs and failures within the studied cohort; they do not estimate prevalence across arbitrary checkers or establish a rule predicting when internal evidence will help.

%% file: sections/08-related.tex
\subhead{Static Analysis and Checker Languages.}
Static-analysis systems provide the substrate that \toolname refines against.
Path-sensitive analyzers, abstract interpreters, and symbolic engines expose guards, states, paths, and diagnostics reusable as evidence~\cite{engler2001bugs,ball2002slam,cousot2005astree,cadar2008klee} and deploy at industrial scale~\cite{calcagno2015infer,bessey2010coverity,sadowski2015tricorder}.
Declarative languages offer a complementary approach by querying program facts through relational rules and points-to specifications~\cite{avgustinov2016ql,jordan2016souffle,bravenboer2009doop}, while semantic patching encodes recurring transformations over systems code~\cite{padioleau2008coccinelle}.
\toolname introduces no new analyzer or query language.
It treats generated checker code as the artifact to be refined.

\subhead{Vulnerability Detection and Patch Evidence.}
Static vulnerability detection has long used taint and data-flow reasoning~\cite{livshits2005finding,jovanovic2006pixy,arzt2014flowdroid}, graph and AST representations~\cite{yamaguchi2014cpg,yamaguchi2012vex}, and learning-based token-, graph-, slice-, and line-level models~\cite{russell2018vulnerability,li2018vuldeepecker,li2021sysevr,fu2022linevul}, though datasets and evaluations show that semantic robustness remains difficult~\cite{fan2020bigvul,chakraborty2021deep}.
Clone- and patch-driven systems use historical fixes as compact witnesses~\cite{jang2012redebug,li2016vulpecker,kim2017vuddy}.
These motivate our use of patches to scope the refinement task.

\subhead{LLM-Based Checker Synthesis.}
Code models and repository agents support program generation and analysis~\cite{chen2021codex,guo2025repoaudit}; LLMDFA decomposes data-flow reasoning and synthesizes tools~\cite{wang2024llmdfa}.
Security studies show why generated code requires validation~\cite{pearce2022copilot,perry2023users,khoury2023chatgpt}.
KNighter~\cite{yang2025knighter} is the closest workflow: it generates, compiles, edits and revalidates checkers using patch and report context.
QLCoder~\cite{wang2025qlcoder} synthesizes CodeQL queries from CVE metadata via an agentic loop with execution feedback and structured tool access.
\toolname adds bounded compiler and checker observations to that setting and attempts warning recovery on silent starts.

\subhead{Feedback-Guided Repair.}
The loop also relates to repair and synthesis under compiler, test, verifier or learned feedback~\cite{clarke2000cegar,solarlezama2006sketch,legoues2012genprog,kim2013par,long2015spr,nguyen2013semfix,mechtaev2016angelix,long2016prophet}, and to LLM work on iterative feedback, tool use and search~\cite{madaan2023selfrefine,shinn2023reflexion,yao2023react,schick2023toolformer,yao2023tree}.
Weak validation proxies and repair overfitting are established concerns~\cite{qi2015plausibility,smith2015overfitting}.
Our follow-up examines their consequences for edited static checkers, where a warning can return at the expected line while its object, scope or path remains wrong.

\subhead{Analyzer Context and Alert Triage.}
LLift uses analyzer reports and post-constraints~\cite{li2024llift}; Chapman et al. interleave analyzer intermediates and prompting for error specifications~\cite{chapman2024interleaving}; IRIS infers taint specifications with CodeQL~\cite{li2025iris}.
LLM4PFA reasons about report feasibility~\cite{du2025llm4pfa}, and ZeroFalse enriches reports before adjudication~\cite{iranmanesh2025zerofalse}.
VulGenie combines patch dependencies, attack and defense flows, and API-rule synthesis~\cite{chen2026vulgenie}.
These studies already use analyzer context.
\toolname instead studies how such context changes an existing checker's implementation and the reports it produces, retaining both the useful edits and their remaining semantic errors.

%% file: sections/09-conclusion.tex
Static-analysis frameworks already compute rich semantic information about how checkers interact with program representations, yet existing LLM-based refinement loops leave this information unused.
\toolname bridges this gap by feeding analyzer-internal evidence into the editing loop, enabling the model to understand why a checker fails rather than only that it does.
Evaluation on KNighter-derived Linux checkers shows that this approach recovers previously silent vulnerable revisions and improves paired discrimination over report-driven refinement alone.
A qualitative follow-up traces each warning-count change to specific checker repairs, such as recognizing lowered callbacks and unwrapping expression casts, confirming that the evidence guides semantically meaningful edits.
Extending this approach to verify target-level semantic correctness beyond warning presence is a promising direction for future work.